\documentclass[prc,superscriptaddress,superscriptaddress,nofootinbib,%
twocolumn,showpacs,preprintnumbers,amsmath,amssymb]{revtex4-1}
\usepackage{graphicx}
\usepackage[breaklinks, colorlinks=true, pdfstartview=FitV]{hyperref}

\newcommand{\Nc}{N_{\text{c}}}
\newcommand{\MeV}{\;\text{MeV}}
\newcommand{\GeV}{\;\text{GeV}}
\newcommand{\muB}{\mu_{\text{B}}}
\newcommand{\mn}{m_{\text{N}}}
\newcommand{\snn}{\sqrt{s_{_{NN}}}}
\newcommand{\bp}{\boldsymbol{p}}
\newcommand{\calN}{\mathcal{N}}
\newcommand{\calchi}{\mathcal{X}}

\begin{document}

\title{Hadron resonance gas and mean-field nuclear matter for baryon
       number fluctuations}
\author{Kenji Fukushima}
\affiliation{Department of Physics, The University of Tokyo,
             7-3-1 Hongo, Bunkyo-ku, Tokyo 113-0033, Japan}

\begin{abstract}
 I give an estimate for the skewness and the kurtosis of the baryon
 number distribution in two representative models; i.e., models of a
 hadron resonance gas and relativistic mean-field nuclear matter.  I
 emphasize formal similarity between these two descriptions.  The
 hadron resonance gas leads to a deviation from the Skellam
 distribution if quantum statistical correlation is taken into account
 at high baryon density, but this effect is not strong enough to
 explain fluctuation data seen in the beam-energy scan at
 RHIC/STAR.\ \ In the calculation of mean-field nuclear matter the
 density correlation with the vector $\omega$-field rather than the
 effective mass with the scalar $\sigma$-field renders the kurtosis
 suppressed at higher baryon density so as to account for the
 experimentally observed behavior of the kurtosis.  We finally discuss
 the difference between the baryon number and the proton number
 fluctuations from correlation effects in isospin space.  The
 numerical results suggest that such effects are only minor even in
 the case of complete randomization of isospin.
\end{abstract}
\maketitle

\section{Introduction}

The phase diagram of matter described by quantum chromodynamics (QCD)
in terms of quarks and gluons, i.e., the QCD phase diagram has not
been unveiled yet in spite of tremendous theoretical and experimental
efforts~\cite{BraunMunzinger:2008tz,Fukushima:2010bq}.  The severest
obstacle lies in the notorious sign problem which prevents the
first-principle lattice QCD simulation from working at high baryon
density, though there are steady progresses to circumvent
it~\cite{Aarts:2013bla}.  There are so many theoretical
speculations on the QCD phase structures but it is next to impossible
to constrain them enough to pin the right one down or to eliminate
unphysical ones.  Even if there were a way to evade the sign problem,
it would still be a highly non-trivial question whether the numerical
simulation can correctly identify the genuine ground state if it
includes a possibility of spatial modulation~\cite{Fukuda:2013ada}.
Taking the continuum limit and overcoming the discretization error
should be crucial to resolve intricate structures such as the critical
point~\cite{Stephanov:1998dy} (see also Ref.~\cite{Fukushima:2008is}
for a heuristic argument) and, if any, the crystalline
condensates~\cite{Nakano:2004cd,Nickel:2009ke} (see
Ref.~\cite{Fukushima:2012mz} for an argument parallel to
Ref.~\cite{Fukushima:2008is} and also Ref.~\cite{Buballa:2014tba} for
a comprehensive review).

It is thus our hope that the experimental data should be able to
constrain diverse candidates of the QCD phase diagram, so that we can
identify the correct answer.  Now that there are reasonable evidences
for the formation of a new state of matter out of quarks and gluons,
that is called the quark-gluon plasma, at high enough energy, some of
future heavy-ion collision programs are directed toward higher baryon
density with lower collision energies.  Such a project to explore the
QCD phase diagram by tuning the collision energy is often called the
beam-energy scan (BES) and the STAR Collaboration at Relativistic
Heavy Ion Collider (RHIC) already published the first BES (i.e.,
BES-I) results~\cite{Adamczyk:2013dal}.  The primary mission of the
BES was to discover the so-called QCD critical point by looking at
fluctuations of conserved quantities such as the baryon number and the
strangeness~\cite{Stephanov:1998dy,Stephanov:2008qz,Asakawa:2009aj}.

So far, there is no appreciable indication that signals for the
critical behavior~\footnote{A new analysis including higher-$p_t$ data
  (that was motivated to improve the statistics) suggests critical
  behavior.  It is still under dispute;  I should note that,
  if $T\sim\Lambda_{\rm QCD}\sim 0.2\GeV$, the kinetic energy should
  be; $p_t^2/2\mn\sim T/2$ leading to $p_t\sim 0.4\GeV$.  It should be
  explained why higher (above $0.8\GeV$) $p_t$ data enhance the
  criticality.  So, in this work, I shall focus on the published data
  only.}, and nevertheless, the BES has turned out to be
extremely intriguing for QCD physics, for our understanding of
finite-density QCD is severely limited and any hint would be useful.
With accumulation of abundant experimental data, it might be even
feasible to find a way for drastic simplification leading to pragmatic
modeling.  We have already witnessed such simplification in RHIC at
high temperature $T$ and low baryon chemical potential $\muB$;  the
statistical thermal
fit~\cite{Cleymans:2005xv,Becattini:2005xt,Andronic:2008gu} and the
hadron resonance gas (HRG) model~(see Ref.~\cite{Borsanyi:2011zz} and
references therein and also Ref.~\cite{Karsch:2010ck} for a recent
study) stunningly reproduce the experimental yields of particles and
they are also consistent with lattice-QCD thermodynamics.  Nobody had
believed in the reality of such an oversimplified description of
non-interacting hadrons before the good agreement to experimental data
was confirmed.  Although the theoretical foundation needs more
investigations, this a bit expedient but profitable tool for data
analysis is as effective for analyzing experimental data taken by the
ALICE Collaboration at Large Hadron Collider (LHC) (see
Ref.~\cite{Stachel:2013zma} and references therein), though minor
deviations were reported.

We cannot, of course, trust the HRG model over the entire QCD phase
diagram away from the chemical freeze-out line.  It is obvious that
the HRG should break down in the region of nuclear matter at low-$T$
and high-$\muB$.  Nuclear physics at $T=0$ has revealed that a
first-order phase transition of liquid-gas (or liquid-vacuum at $T=0$)
should take place at $\muB=M_N-B$ with $M_N$ and $B$ being the nucleon
mass and the binding energy $B\simeq 16\MeV$~\cite{Tatsumi:2011tt}.
Some years ago an interesting possibility was
demonstrated~\cite{Floerchinger:2012xd};  the chemical freeze-out
condition at low-$T$ and high-$\muB$ could be rather sensitive to
nuclear matter properties.  The present work aims to pursue the idea
along the same line to show the agreement for not only the chemical
freeze-out condition but also the fluctuations.

One might have an impression that the HRG is a sort of opposite to
nuclear matter and one should abandon the HRG immediately to switch to
the nuclear physics terrain.  This intuition is not totally correct,
however, and we know that the independent quasi-particle picture makes
good sense inside of nuclei and nuclear matter.  Hence, on the formal
level, the HRG-like model with ``renormalized'' parameters may have a
chance to work continuously from low-$\muB$ to high-$\muB$.  Indeed,
the relativistic mean-field (RMF) model of nuclear matter is designed
in this spirit.  The simplest setup of the RMF is the
$\sigma$-$\omega$ model~\cite{Walecka:1974qa} as was adopted in
Ref.~\cite{Floerchinger:2012xd}.  This model deals with nucleons as
relativistic quasi-particles moving in the scalar mean-field $\sigma$
and the vector mean-field $\omega$.  I note that we can safely
neglect $\pi$ fluctuations as long as we concern the baryon number at
small $T$.  If needed, I can extend my present analysis so as to
include $\pi$ fluctuations, for example, with the renormalization
group improvement~\cite{Drews:2013hha}.

This paper is organized as follows:  I give a detailed description of
fluctuations within the framework of the HRG model in
Sec.~\ref{sec:HRG}.  Then, based on the similarity to the HRG model,
I introduce the RMF model in Sec.~\ref{sec:similarity} and I present
my central numerical results from the RMF model in
Sec.~\ref{sec:central}.  In Sec.~\ref{sec:suppression} I give more
considerations on the microscopic structures of my numerical
results.  I also cover discussions on the difference between the
baryon number and the proton number to discover that the diffusion in
isospin space does not affect my results as long as the Boltzmann
approximation makes sense, which is addressed in
Sec.~\ref{sec:isospin}.  I finally summarize this work in
Sec.~\ref{sec:summary}.

\section{Fluctuations and the hadron resonance gas}
\label{sec:HRG}

First of all, before going into the descriptions of the HRG model, I
should elucidate physical observables of my interest.  I follow the
standard convention as used in Ref.~\cite{Karsch:2010ck} for thermal
fluctuations which are derived from the derivatives of the pressure
with respect to the relevant chemical potentials.  For the baryon
number fluctuation, thus, I calculate the following dimensionless
quantities:
\begin{equation}
 \chi_B^{(n)} \equiv \frac{\partial^n}{\partial(\muB/T)^n}
  \frac{p}{T^4}\;,
\label{eq:chi}
\end{equation}
from which I can construct the mean value (i.e., the particle
number); $M\equiv VT^3\chi_B^{(1)}$.  For an arbitrary distribution I
can define the Gaussian width $\sigma^2$ together with the
non-Gaussian fluctuations such as the skewness $S$ and the kurtosis
$\kappa$ as~\cite{Stephanov:2008qz,Karsch:2010ck}:
\begin{equation}
 \frac{\sigma^2}{M} \equiv \frac{\chi_B^{(2)}}{\chi_B^{(1)}} \;,\quad
 S\sigma \equiv \frac{\chi_B^{(3)}}{\chi_B^{(2)}} \;,\quad
 \kappa\sigma^2 \equiv \frac{\chi_B^{(4)}}{\chi_B^{(2)}} \;.
\label{eq:fluc}
\end{equation}
Therefore, once some theoretical estimates provide us with the
pressure $p$ as a function of $\muB$, I can give a prediction for
these fluctuations under an assumption of the dominance of thermal
fluctuations.

Second, to make a contact with the collision experiment, it is
necessary to relate the collision energy $\snn$ and $T$ and $\muB$.
Fortunately, such parametrization of $T(\snn)$ and $\muB(\snn)$ has
been well established along the chemical freeze-out
line~\cite{Cleymans:2005xv} that reads:
\begin{align}
 & T(\muB) = a - b\,\muB^2 - c\,\muB^4 \;,
\label{eq:freezeT}\\
 & \muB(\snn) = \frac{d}{1+e\snn} \;,
\label{eq:freezemu}
\end{align}
where parameters are chosen as $a=0.166\GeV$, $b=0.139\GeV^{-1}$,
$c=0.053\GeV^{-3}$, $d=1.308\GeV$, and $e=0.273\GeV^{-1}$ to reproduce
experimentally observed particle yields.  Charge and strangeness
chemical potentials, $\mu_Q$ and $\mu_S$, are also parametrized in a
similar manner.  In my present analysis, I numerically checked that
the inclusion of $\mu_Q$ and $\mu_S$ hardly changes the fluctuation
results, and so I neglect them for clarity of presentation.  These
definitions and parametrizations are robust and unchanged for any
model applications.

Now I take a step toward the HRG model.  Let us start with a simple
demonstration of free nucleon gas and then proceed to the realistic
HRG model next.  In the estimate with non-interacting hadrons (in
which the canonical factor $\gamma$ is not included) I make use of
the standard expression of the free grand canonical partition
function.  That is, the pressure from baryons (fermions) is prescribed
as
\begin{align}
 p_{\text{free}}(\mn,\muB) &= \sum_i^N 2T\int\frac{d^3 p}{(2\pi)^3}\,
  \Bigl\{ \ln\bigl[ 1\!+\!e^{-(\varepsilon_p \!-\! \muB)/T} \bigr]
  \notag\\
  &\qquad\qquad + \ln\bigl[ 1+e^{-(\varepsilon_p+\muB)/T} \bigr]
  \Bigr\} \;.
\label{eq:p_free}
\end{align}
Here $N$ is $2$ for nucleons corresponding to the isospin degeneracy
and the pressure depends on the nucleon mass $\mn$ through the energy
dispersion relation: $\varepsilon_p\equiv\sqrt{\bp^2+\mn^2}$.  I can
then take the derivatives of the above expression, which results in
\begin{equation}
 \chi_B^{(n)} = \frac{4}{T^3}\int\frac{d^3 p}{(2\pi)^3}\,
  X^{(n)}(p) \;,
\end{equation}
where $4$ appears from the spin and the isospin degeneracy (for $N=2$)
and the integrands read:
\begin{align}
 X^{(1)} &= n_p - \bar{n}_p\;,\notag\\
 X^{(2)} &= n_p(1-n_p) + \bar{n}_p(1-\bar{n}_p)\;,\notag\\
 X^{(3)} &= n_p(1\!-\!n_p)(1\!-\!2n_p)
  - \bar{n}_p(1\!-\!\bar{n}_p)(1\!-\!2\bar{n}_p)\;,
\label{eq:X}\\
 X^{(4)} &= (1-6n_p+6n_p^2) n_p(1-n_p) \notag\\
        &\qquad\qquad\qquad
           + (1-6\bar{n}_p+6\bar{n}_p^2)\bar{n}_p(1-\bar{n}_p) \notag
\end{align}
with $n_p\equiv[e^{(\varepsilon_p-\muB)/T}+1]^{-1}$ and
$\bar{n}_p\equiv[e^{(\varepsilon_p+\muB)/T}+1]^{-1}$ being the
Fermi-Dirac distribution functions for nucleons and anti-nucleons.  I
can continue taking the derivatives for even larger $n$ if needed.

\begin{figure}
 \includegraphics[width=\columnwidth]{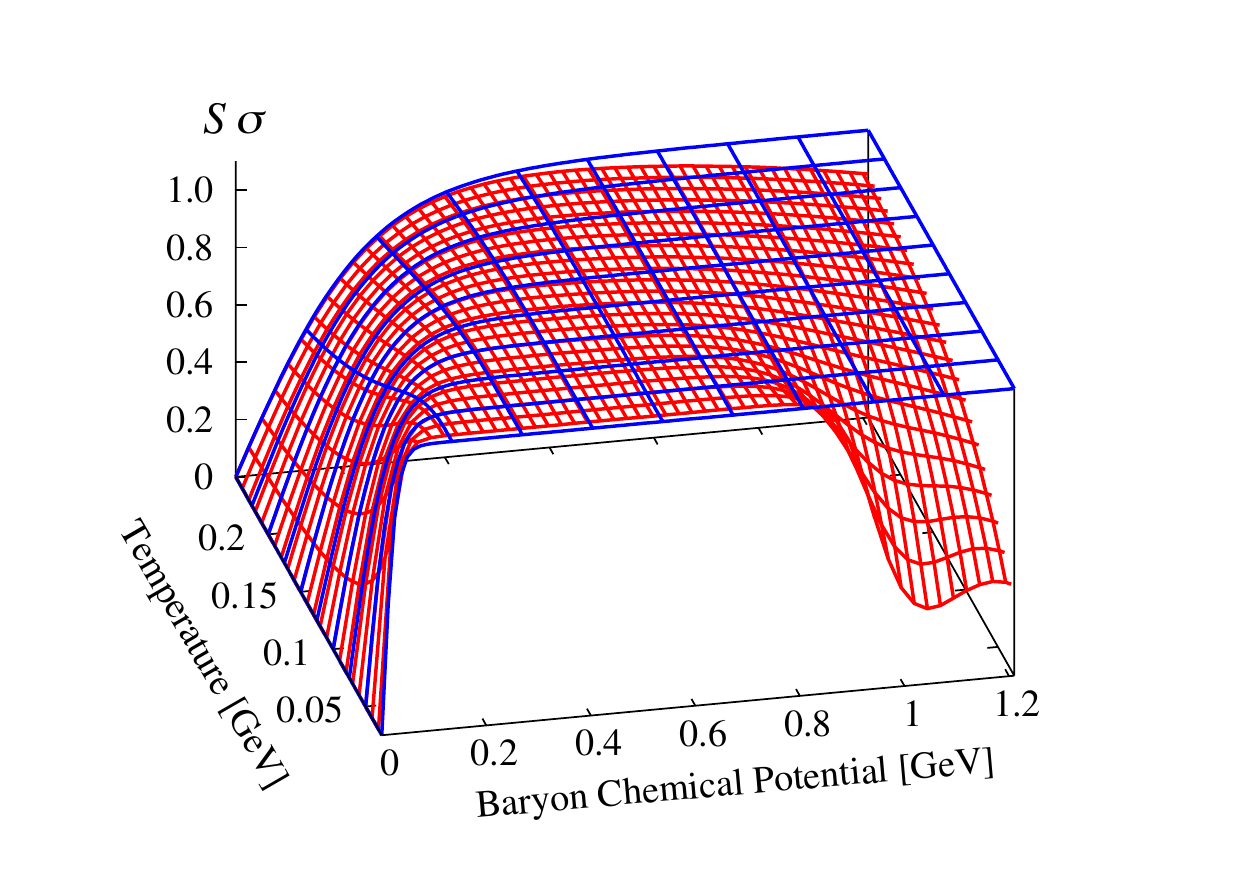}
 \caption{(Color online) Skewness of the baryon number estimated in
   the HRG (THERMUS2.3) by the (red) fine mesh.  The (blue) sparse
   mesh represents the Skellam expectation: $\tanh(\muB/T)$.}
 \label{fig:HRGS}
\
\includegraphics[width=\columnwidth]{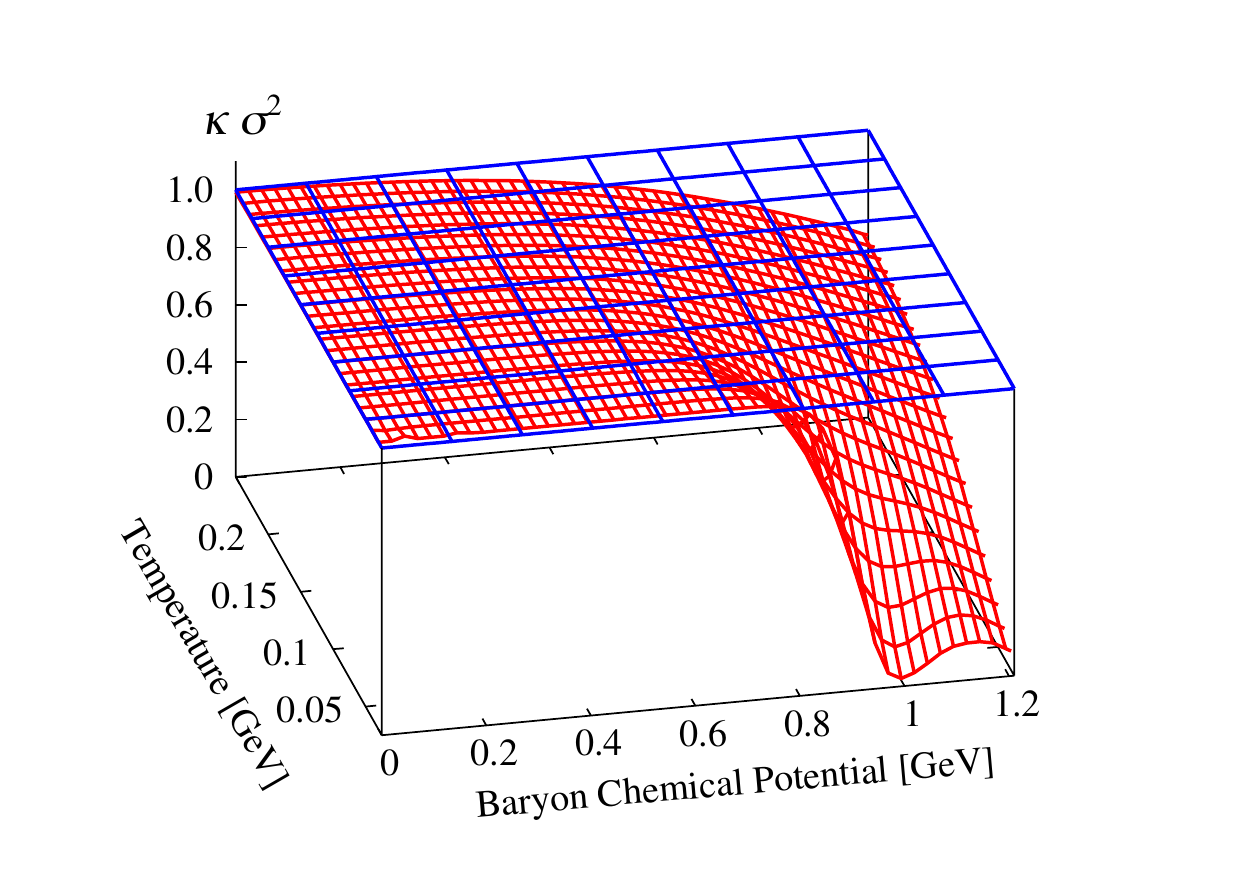}
 \caption{(Color online) Kurtosis of the baryon number estimated in
   the HRG (THERMUS2.3) by the (red) fine mesh.  The (blue) sparse
   mesh represents the Skellam expectation that is the unity.}
 \label{fig:HRGk}
\end{figure}

In the Boltzmann approximation that is valid when $n_p$ and
$\bar{n}_p$ are both dilute, I can neglect the quantum statistical
factors of non-linear $n_p$ and $\bar{n}_p$ terms.  Then, I can
approximate Eq.~\eqref{eq:X} as
$X^{(2)}\approx X^{(4)}\approx (e^{\muB/T}+e^{-\muB/T})
e^{-\varepsilon_p/T}$ and
$X^{(3)}\approx (e^{\muB/T}-e^{-\muB/T}) e^{-\varepsilon_p/T}$.  In
this particular limit I can readily derive:
\begin{equation}
 S\sigma = \tanh(\muB/T)\;,\qquad
 \kappa\sigma^2 = 1\;,
\label{eq:skellam}
\end{equation}
which are nothing but the Skellam expectations.  I can easily
generalize the above derivation of Eq.~\eqref{eq:skellam} to a
superposition of arbitrary $N$ with different masses to find that
Eq.~\eqref{eq:skellam} still holds after all.  This is because
$e^{\muB/T}\pm e^{-\muB/T}$ is always factored out and the remaining
integrand is common for $X^{(2)}$, $X^{(3)}$, and $X^{(4)}$.

Let us then quantify the breakdown of the Boltzmann approximation
explicitly by scanning the 3D landscape of $S\sigma$ and
$\kappa\sigma^2$ for various $T$ and $\muB$.  In Figs.~\ref{fig:HRGS}
and \ref{fig:HRGk} we show our results from (not a free nucleon gas
but) the HRG model using the particle data contained in the THERMUS2.3
package (by red fine mesh) as well as the Skellam predictions (by blue
sparse mesh).  It is clear from the figures that the quantum
correlation certainly suppresses both $S\sigma$ and $\kappa\sigma^2$
in the high-density region where $n_p$ is not really dilute.  I
should note that the HRG model can describe the onset behavior of
finite baryon density but does not have dynamics enough to realize a
first-order liquid-gas phase transition of nuclear matter (and this is
why I do not show HRG results at temperatures smaller than a few tens
MeV in Figs.~\ref{fig:HRGS} and \ref{fig:HRGk}).  Although this
suppression effect is noticeable along the chemical freeze-out line as
in Figs.~\ref{fig:skewness} and \ref{fig:kurtosis}, it is not
sufficiently strong for reproducing the trend of the experimental
data.  In short, the quantum correlation is weak, as correctly
speculated in Ref.~\cite{Karsch:2010ck}, because the baryon density
never gets large enough on the chemical freeze-out line.

\begin{figure}
 \includegraphics[width=\columnwidth]{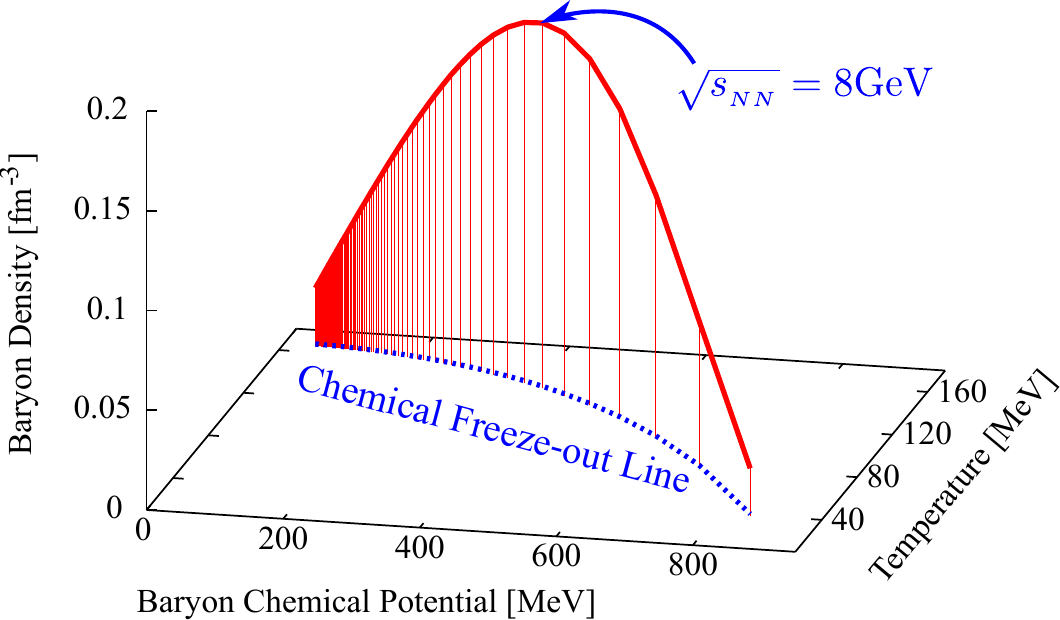}
 \caption{(Color online) HRG-estimated baryon density (including not
   only nucleons but all baryonic resonances of the particle data
   contained in the THERMUS2.3 package) as a function of $T$ and
   $\muB$.  The nucleon contribution is nearly a half of shown
   results.  The vertical lines represent the collision energy with
   spacing by $1\GeV$.  The extremal point corresponds to $\snn\simeq
   8\GeV$.  The chemical freeze-out line is drawn according to
   Eqs.~\eqref{eq:freezeT} and \eqref{eq:freezemu}.}
 \label{fig:density}
\end{figure}

To have a feeling about how the baryon density behaves on the chemical
freeze-out line, I shall make a plot of the integrated baryon density
in the standard unit of $\text{fm}^{-3}$ in Fig.~\ref{fig:density}.
The vertical thin lines correspond to the collision energy $\snn$ with
spacing by $1\GeV$.  The lowest collision energy in
Fig.~\ref{fig:density} starts with $\snn=2\GeV$, and the maximum of
the baryon density is found at $\snn\sim 8\GeV$.  It is
interesting that this maximum position precisely coincides with the
triple-point-like region as speculated in
Ref.~\cite{Andronic:2009gj}.  This coincidence is not accidental;  in
Ref.~\cite{Andronic:2009gj} the triple-point-like region was
recognized based on the horn structure in $K^+/\pi^+$ that is
sensitive to the strangeness chemical potential; $\mu_S$.  If the bulk
system maintains zero strangeness, it is not hard to confirm that
$\mu_S$ is almost proportional to $\muB$ within an effective model
framework~\cite{Fukushima:2009dx}.  In this way, naturally,
$K^+/\pi^+$, $\Lambda/\pi^-$, $\Xi/\pi^-$, etc have a peak structure
at $\snn\simeq 8\GeV$ with which the baryon density is maximized.

As a final related remark I point out that the effect of the
strangeness and the charge conservation is only of a few percent order
in $S\sigma$ and $\kappa\sigma^2$ along the chemical freeze-out line.
I have checked this numerically by adopting $\mu_Q$ and $\mu_S$
parametrized along the chemical freeze-out line~\cite{Karsch:2010ck}.
I then observed that $S\sigma$ and $\kappa\sigma^2$ in
Figs.~\ref{fig:skewness} and \ref{fig:kurtosis} are pushed down by a
few percent at most as compared to the current $\mu_Q=\mu_S=0$ case.
This check justifies my discussions without $\mu_Q$ and $\mu_S$ taken
into account.

\section{Similarity between HRG and RMF}
\label{sec:similarity}

It is nuclear matter (that is a self-bound system of infinite
nucleons) that lies in the opposite limit to the non-interacting
matter described by the HRG model.  Nevertheless, theoretically
speaking, the formulation of nuclear matter, namely the RMF, is not
such far from the HRG model or they actually share similarity to
some extent.

The simplest RMF is known as the $\sigma$-$\omega$ model defined by
the partition function:
\begin{align}
 p &= 2\cdot 2T\int\frac{d^3p}{(2\pi)^3} \Bigl\{ \ln\bigl[
  1+e^{-(\varepsilon_p-\muB^\ast)/T} \bigr] \notag\\
 &\qquad + \ln\bigl[ 1+e^{-(\varepsilon_p+\muB^\ast)/T} \bigr] \Bigr\}
  - \frac{m_\sigma^2 \sigma^2}{2} + \frac{m_\omega^2 \omega^2}{2} \;,
\label{eq:rmf}
\end{align}
where the quasi-particle dispersion relation is
$\varepsilon_p\equiv\sqrt{\bp^2+\mn^{\ast 2}}$.  Here, quantities with
asterisk are ``in-medium'' or ``renormalized'' ones which contain a
shift by the mean-field as
\begin{equation}
 \mn^\ast \equiv \mn - g_\sigma \sigma\;,\qquad
 \muB^\ast \equiv \muB - g_\omega \omega\;.
\end{equation}
These mean-fields of $\sigma$ and $\omega$, or equivalently,
$\mn^\ast$ and $\muB^\ast$ are determined with the stationary
conditions:
$\partial\Omega/\partial \sigma=\partial\Omega/\partial \omega=0$,
which lead to the gap equations.  By choosing the model parameters
appropriately~\cite{Buballa:1996tm}; i.e., $\mn=939\MeV$,
$m_\sigma=550\MeV$, $m_\omega=783\MeV$, $g_s=10.3$, $g_\omega=12.7$,
we can reproduce the saturation properties of symmetric nuclear matter
with the saturation density given by
$0.17\;\text{nucleons}/\text{fm}^3$ and the binding energy per nucleon
given by $16.3\MeV$.  I note that this simplest $\sigma$-$\omega$
model fails in reproducing the empirical value of the compressibility
of symmetric nuclear matter~\cite{Blaizot:1980tw}.  It is possible to
overcome this problem by extending the model with self-coupling
potential of the mean-fields.  For the fluctuations of my present
interest, however, such improvement of the model makes only minor
modifications on the final results~\cite{sasaki}.  This also implies
that a different choice of $m_\sigma$, e.g.\ $500\MeV$ would not
change the final results because $g_s$ and $g_\omega$ should be
readjusted to reproduce the saturation density and the binding energy,
and so the difference would be the compressibility only.

From Eq.~\eqref{eq:rmf} it is obvious that the RMF estimate should
reduce to nothing but the HRG estimate or Eq.~\eqref{eq:p_free} if I
freeze the implicit dependence on $\muB$ through the solutions of
$\sigma$ and $\omega$, or equivalently, $\mn^\ast$ and $\muB^\ast$.
In this sense we can interpret the RMF treatment as a variation of the
HRG model augmented with mean-fields.  Unlike the HRG model, however,
the mean-fields have implicit dependence on $\muB$, from which I
should anticipate non-trivial contributions for the fluctuations.

In closing of this section, I make an explicit statement about the
validity regions of the HRG model and the RMF models.  The HRG model
is the most successful at the top energy of the RHIC, but the
agreement of the thermal model fit to the experimental data slightly
becomes worse for the LHC data.  There is no clear explanation for
this, but it is conceivable that the HRG model works the best near the
crossover region of deconfinement.  The meson sector of the HRG model
is a valid picture in a fictitious world of $\Nc\to\infty$ with which
meson interactions would be turned off.  The baryon sector behaves
differently, however, and so the HRG model should naturally break down
at high baryon density.  A conservative estimate for this would
suggest a validity region, $\muB<T$, that corresponds to
$\snn\gtrsim 10\GeV$.  On the other hand, the RMF model is supposed to
describe nuclear matter which is reached at small $\snn$, and the
validity region is limited by my approximation to neglect pion
fluctuations.  Although the effect of pions is indirect for the baryon
number fluctuations, it could make a quantitative modification if the
temperature is comparable to the pion mass.  This condition would
translate into the validity region $\snn\lesssim 10\GeV$ in the energy
unit.  So, one may well expect that the HRG model at high energy should
be taken over smoothly by the RMF model in the intermediate energy
$\snn\sim 10\GeV$, which could be of course understood as another
manifestation of the triple-point-like region~\cite{Andronic:2009gj}.

\section{Central numerical results}
\label{sec:central}

\begin{figure}
 \includegraphics[width=\columnwidth]{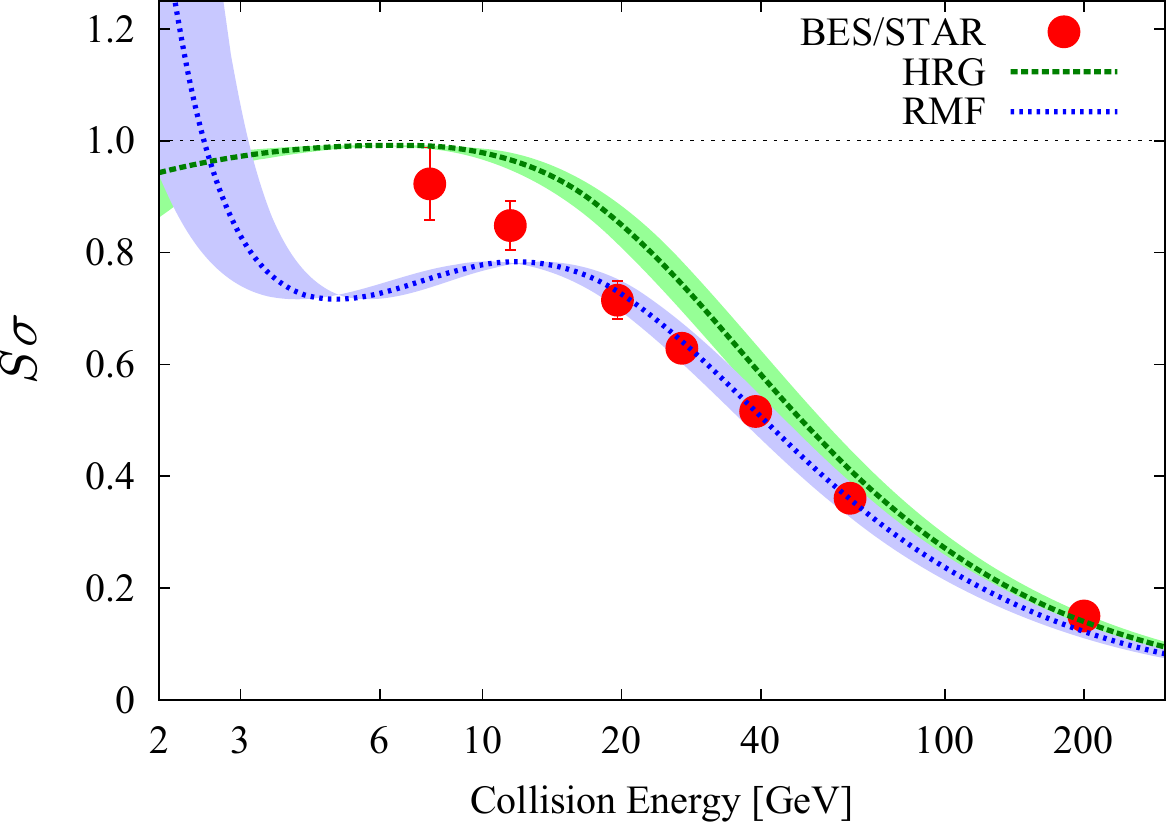}
 \caption{(Color online) Skewness of the baryon number distribution.
   The red dot, the green dotted line, and the blue dashed line
   represent the results from the BES/STAR, the HRG, and the RMF,
   respectively.  The bands represent uncertainty from the freeze-out
   $\muB$ by $\pm10\%$.}
 \label{fig:skewness}
\end{figure}

\begin{figure}
 \includegraphics[width=\columnwidth]{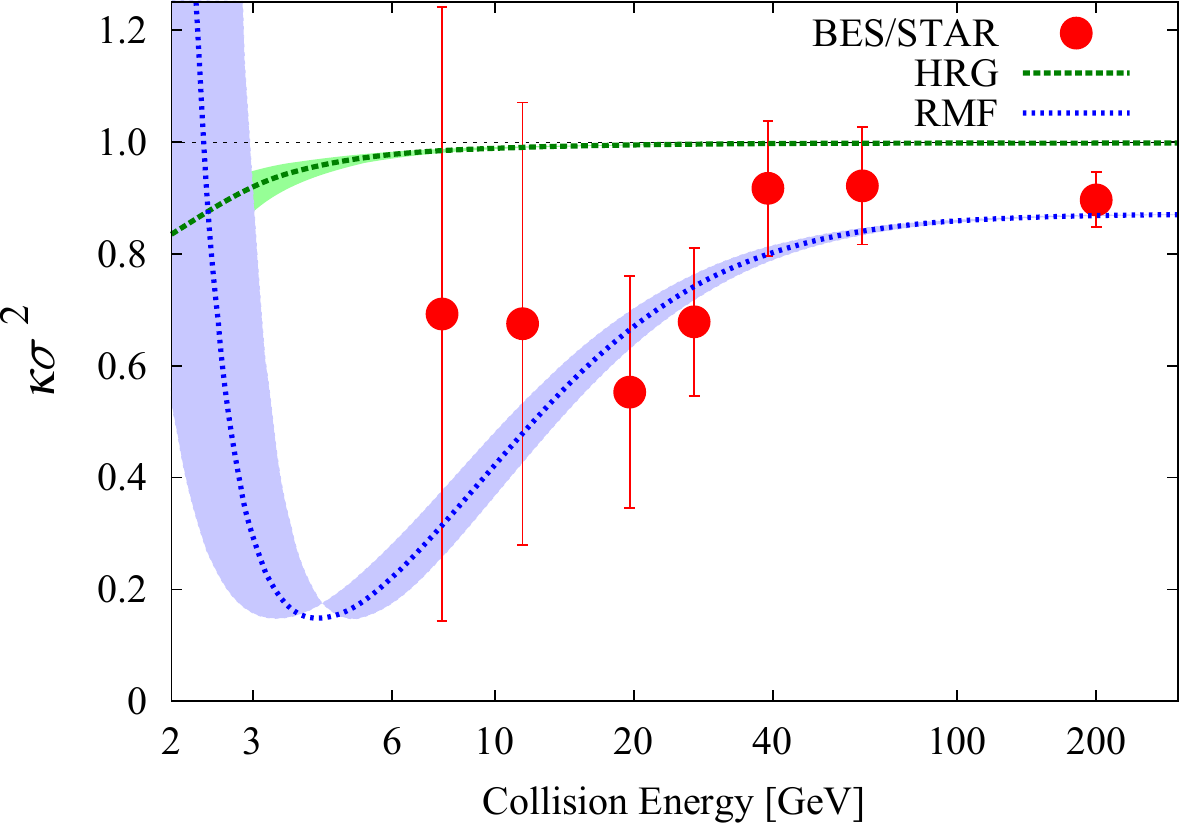}
 \caption{(Color online) Kurtosis of the baryon number distribution.
   The legend convention is the same as in Fig.~\ref{fig:skewness}.
   The bands represent uncertainty from the freeze-out $\muB$ by
   $\pm10\%$.}
 \label{fig:kurtosis}
\end{figure}

Figures~\ref{fig:skewness} and \ref{fig:kurtosis} show my results for
$S\sigma$ and $\kappa\sigma^2$ estimated in the HRG (green dotted
line) and in the RMF (blue dashed line) on top of the BES/STAR data
(red dots).  I note that the parametrization of the chemical
freeze-out line, \eqref{eq:freezeT} and \eqref{eq:freezemu}, may have
some uncertainty particularly at small $\snn$.  To quantify the
sensitivity I varied $\muB$ by $\pm10\%$ to add the band on each line
in Figs.~\ref{fig:skewness} and \ref{fig:kurtosis}.  Now let me
briefly discuss particular two among the non-trivial features
noticeable in these figures.

One is that the HRG model may have a richer structure than the Skellam
distribution.  Actually it was clearly stated in
Ref.~\cite{Karsch:2010ck} that the Skellam predictions come from the
Boltzmann approximation.  If the baryon density gets large, therefore,
one naturally expects modifications on the distribution.  More
specifically, as seen in Figs.~\ref{fig:skewness} and
\ref{fig:kurtosis}, the kurtosis is not necessarily the unity at small
$\snn$.  This effect is not such substantial, but it would be
interesting to reveal how the quantum correlation would affect the
distribution in a wider region away from the chemical freeze-out
line.

The other is that $\kappa\sigma^2$ in the RMF is suppressed at smaller
$\snn$ thus larger $\muB$.  In fact the RMF-estimated $\kappa\sigma^2$
happens to approach the experimental data.  It is, of course, the
interaction effect that modifies $S\sigma$ and $\kappa\sigma^2$.
Then, an immediate question that comes to my mind is which of
$\sigma$ and $\omega$ should be more responsible for the suppression
seen in Fig.~\ref{fig:kurtosis}.  One may well consider that the
in-medium effective mass can bring about the leading effect of the
interactions, which is indeed the case whenever the Hartree
approximation works.  In the present problem, as we will see in the
next section, the situation is rather involved.  Because I take the
$\muB$ derivatives to compute the baryon fluctuations, it turns out to
be $\muB^\ast$ and thus $\omega$ that play the essential role for
forming a peculiar shape of $\kappa\sigma^2$ in
Fig.~\ref{fig:kurtosis}.  Therefore, my study, as I will explain
later, brings me a conclusion that the renormalization of $\muB$
caused by $\omega$ suppresses $\kappa\sigma^2$, while the in-medium
mass coupled with $\sigma$ does the opposite.  I comment that, in
view of Figs.~\ref{fig:skewness} and \ref{fig:kurtosis}, the
fluctuations grow up again when $\snn$ reaches below $\sim 4\GeV$.
This low-$\snn$ enhancement of the fluctuations is simply because of
the criticality when the chemical freeze-out line hits the liquid-gas
critical point of nuclear matter~\cite{Chomaz:2004nw} that is located
at $T\simeq 21\MeV$ and $\muB\simeq 906\MeV$ in my RMF setup.

\section{What causes the suppression?}
\label{sec:suppression}

As I mentioned previously, if I fix $\mn^\ast$ and $\muB^\ast$ at
the vacuum values; i.e., $\mn$ and $\muB$, and then take the $\muB$
derivatives, the results for $S\sigma$ and $\kappa\sigma^2$ are
identical to what is referred to by the HRG in
Figs.~\ref{fig:skewness} and \ref{fig:kurtosis}, which I have
numerically checked.  They are not exactly the same because the
genuine HRG results have contributions also from higher baryonic
resonances.

If I include the in-medium mass effect only, the $\muB$ derivative
hits the implicit dependence in $n_p$ and $\bar{n}_p$ and, for
example, the first derivative reads:
\begin{equation}
 \frac{\partial n_p}{\partial(\muB/T)}
  = (1-\varepsilon_p') n_p(1-n_p) \simeq
  (1-\varepsilon_p') n_p
\end{equation}
in the Boltzmann approximation.  Here $\varepsilon_p'$ represents
$\partial\varepsilon_p/\partial\muB$.  I find a similar expression
for $\bar{n}_p$ with an overall minus sign and with $-\varepsilon_p'$
changed to $+\varepsilon_p'$.

At high density I can neglect the anti-particle contribution from
$\bar{n}_p$, and moreover, $\varepsilon_p'$ is negative because the
effective mass $\mn^\ast$ generally decreases with increasing density.
This means that $\partial n_p/\partial(\muB/T)$ is \textit{greater}
than $n_p$ by an enhancement factor $1-\varepsilon_p'>1$.  In the
approximation to neglect higher derivatives in terms of $\muB$,
therefore, $S\sigma$ and $\kappa\sigma^2$ should get larger,
respectively, by $(1-\varepsilon_p')^3$ and $(1-\varepsilon_p')^4$.

In contrast to this behavior of $\mn^\ast$, the effect of the
renormalized chemical potential $\muB^\ast$ yields a suppression
factor by $\partial\muB^\ast/\partial\muB
=1-g_\omega(\partial\omega/\partial\muB)$
where $\omega$ is proportional to the baryon density, so that I can
conclude that $\partial\omega/\partial\muB>0$.  The above-mentioned
arguments have been carefully confirmed in my numerical
calculations.

\begin{figure}
 \includegraphics[width=\columnwidth]{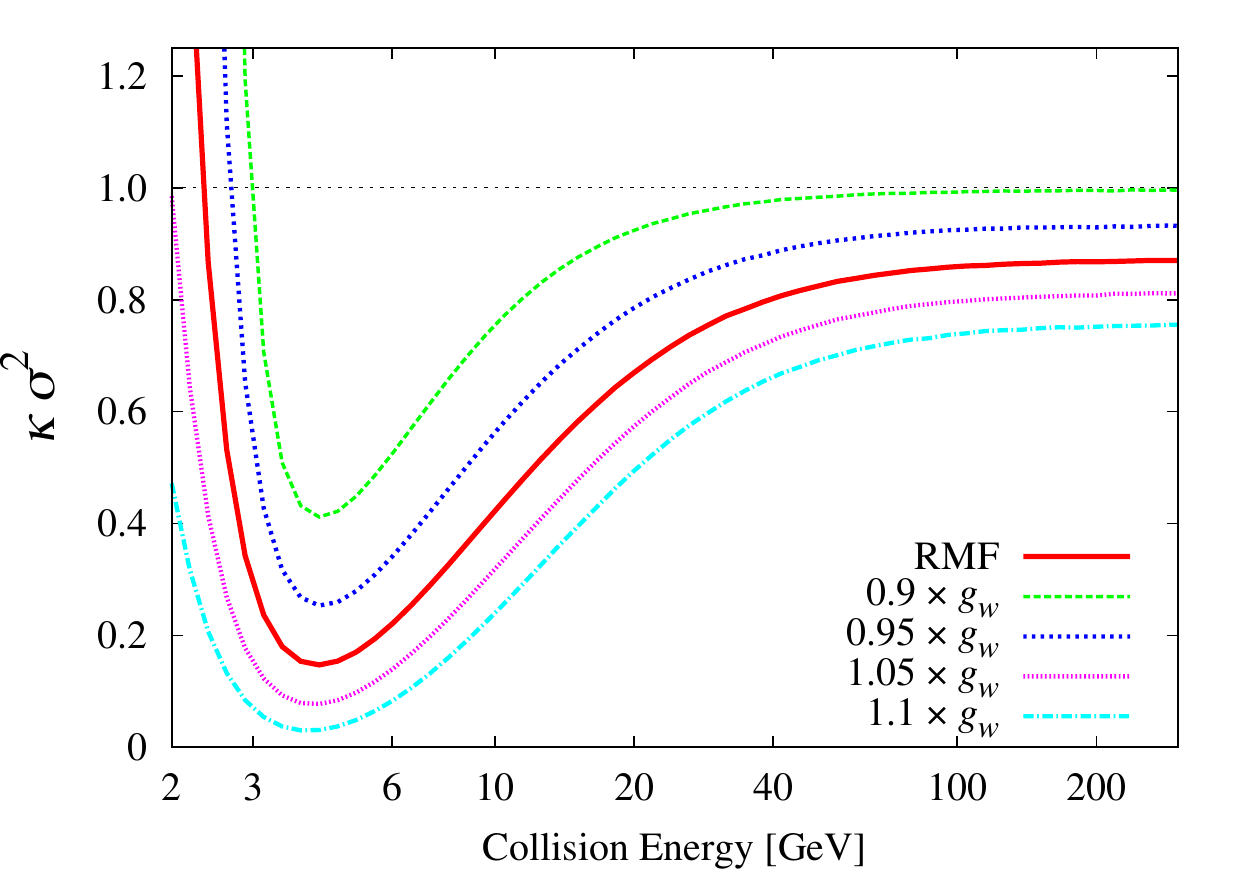}
 \caption{(Color online) Kurtosis calculated in the RMF for various
   vector couplings.}
 \label{fig:kurtosis_test}
\end{figure}

Let us see the numerical check from a different view point.  I change
the strength of the vector coupling $g_\omega$ by hand to find that
$\kappa\sigma^2$ is certainly modified in a way consistent with the
above qualitative arguments, as is transparent in
Fig.~\ref{fig:kurtosis_test};  the entire curve goes down for larger
$g_\omega$.  I should note, however, that I cannot infer $g_\omega$ from
a fit of the model results to the experimental data. This is because
I simply vary $g_\omega$ not adjusting other parameters to reproduce
the saturation properties of nuclear matter.  In this sense, thus, my
results in Fig.~\ref{fig:kurtosis_test} should not be regarded as
anything beyond a test purpose.

\section{Effects of isospin correlations}
\label{sec:isospin}

So far, I have discussed a quantitative comparison assuming that the
experimentally measurable quantities of the proton number fluctuations
are somehow to be identified as the baryon number fluctuations.  One
may have wondered if it really works or not.  In fact such
identification requires a non-trivial assumption about independence
between neutrons and protons as is the case in the HRG calculation.
One can readily understand this by expanding higher powers of
$N_B=N_p+N_n$ where $N_p$ and $N_n$ are, respectively, the (net)
proton number and the (net) neutron number.  For the simplest example,
the quadratic fluctuation consists of
\begin{equation}
 \chi_B^{(2)} = \frac{1}{VT^3}\bigl(\langle N_B^2 \rangle
  - \langle N_B \rangle^2 \bigr)
  = \chi_p^{(2)} + \chi_n^{(2)} + 2\chi_{pn}^{(2)} \;,
\label{eq:quadratic}
\end{equation}
where
\begin{equation}
 \chi_{pn}^{(2)} \equiv \frac{1}{VT^3}\bigl(
  \langle N_p N_n \rangle \!-\! \langle N_p\rangle \langle N_n\rangle
   \bigr) \;.
\end{equation}
If the proton and the neutron behave independently from their isospin
partners, there is no connected contribution in the correlation
function of $N_p$ and $N_n$; i.e., 
$\langle N_p N_n\rangle=\langle N_p\rangle\langle N_n\rangle$ and the
last term involving $\chi_{pn}^{(2)}$ in Eq.~\eqref{eq:quadratic}
vanishes.  As long as I do not consider isospin symmetry violation,
the neutron fluctuation should be just identical with the proton
fluctuation, so that I can conclude $\chi_B^{(2)} = 2\chi_p^{(2)}$
immediately from Eq.~\eqref{eq:quadratic}.  I can continue similar
arguments to deduce that $\chi_B^{(n)} = 2\chi_p^{(n)}$ in general.
Therefore, obviously, this factor 2 is canceled out in the
dimensionless ratios and $S\sigma$ and $\kappa\sigma^2$ of protons
take the same value as those of baryons (nucleons).

This argument is valid as long as I consider a free gas of baryons
only.  It is known, however, that off-diagonal components of the
susceptibility such as $\chi_{ud}\propto \chi_{pn}^{(2)}$ are
non-vanishing as observed in the lattice-QCD
simulation~\cite{Gavai:2005yk} as well as in the model
studies~\cite{Mukherjee:2006hq,Cristoforetti:2010sn}.  I do not go
into technical details here but simply note that non-zero $\chi_{ud}$
is induced by different behavior of the Polyakov loop and the
anti-Polyakov loop in a finite-density environment described by the
Polyakov-loop extended Nambu--Jona-Lasinio
model~\cite{Fukushima:2003fw,Megias:2004hj}.  Physically speaking,
different flavors communicate to each other through confining gluons
to form pions.  It is important to mention that $\chi_{ud}$ itself is
finite also in the HRG calculation, which is attributed to pions
rather than baryons.  Then, a non-zero $\chi_{pn}^{(2)}$ of baryons
should be induced by $\chi_{ud}\neq0$ after all.  Since I cannot
avoid relying on another assumption to give a concrete estimate of
induced $\chi_{pn}^{(2)}$, I shall postpone numerical analyses along
this line into another publication.

Recently a more dynamical origin of isospin correlations has been
discussed in Ref.~\cite{Kitazawa:2012at}.  That is, residual
interactions after the chemical freeze-out can change $p$ into $n$ and
vice versa.  Of course, in the first approximation, I do not have to
think of weak processes because the life time of matter in the
heavy-ion collision is of order of the strong interaction.  Still,
such a mixing between $p\leftrightarrow n$ is allowed by the strong
interaction involving $\pi^0$ and $\pi^-$ through an intermediate
state of $\Delta^+(1232)$ and $\Delta^0(1232)$.  It should be a quite
complicated procedure to establish any reliable evaluation for these
contributions to $\chi_{pn}^{(2)}$, but I can drastically simplify
the theoretical calculation in the limit of complete mixing or
randomization, that is the limit opposite to complete independence in
isospin space.

In this special case of complete randomization of isospin, it is a
natural anticipation to presume that each (anti-) nucleon is either a
(anti-) proton or a (anti-) neutron with equal probability.
Therefore, the distribution of $\calN_p$ is the binomial one with the
mean value given by $\calN_B/2$~\cite{Kitazawa:2012at}, where
$\calN_p$ and $\calN_B$ are not the net quantities but the
\textit{absolute} proton number and the \textit{absolute} baryon
(nucleon) number.  That is, $N_p=\calN_p-\calN_{\bar{p}}$,
$N_B=\calN_B-\calN_{\bar{B}}$, etc.  Thus, for a given $\calN_B$ and
$\calN_{\bar{B}}$ (for which the average is denoted by
$\langle\cdots\rangle_B$), I expect:
\begin{align}
 & \bar{\calN}_p = \langle \calN_p\rangle_B = \frac{1}{2} \calN_B\;,
\label{eq:binom1}\\
 & \langle (\calN_p-\bar{\calN}_p)^2\rangle_B = \frac{1}{4}\calN_B\;,\\
 & \langle (\calN_p-\bar{\calN}_p)^3\rangle_B = 0\;,\\
 & \langle (\calN_p-\bar{\calN}_p)^4\rangle_B
  = \frac{1}{16}\calN_B(3\calN_B-2)\;,
\label{eq:binom4}
\end{align}
and so on according to the binomial distribution.  I note that
Eqs.~\eqref{eq:binom1}-\eqref{eq:binom4} are $T$ independent unlike
the thermal distribution.

I am now ready to express the proton number fluctuations in terms of
baryon ones.  For $n$-th order fluctuation I have:
\begin{equation}
 \chi_p^{(n)} = \frac{1}{VT^3}\biggl\langle\!\!\!\biggl\langle
  \Bigl\langle \Bigl( \calN_p-\calN_{\bar{p}}-
  \Bigl\langle\!\!\Bigl\langle \frac{\calN_B-\calN_{\bar{B}}}{2}
  \Bigr\rangle\!\!\Bigr\rangle \Bigr)^n \Bigr\rangle_B
  \biggr\rangle\!\!\!\biggr\rangle\;,
\end{equation}
where $\langle\!\langle\cdots\rangle\!\rangle$ represents an
average over the distribution of $\calN_B$ and $\calN_{\bar{B}}$.

Using these relations I can easily prove, for example, the following
of the quadratic ($n=2$) fluctuation:
\begin{equation}
 \chi_p^{(2)} = \frac{1}{4}\chi_B^{(2)}
  + \frac{1}{4VT^3} \langle\!\langle \calN_B+\calN_{\bar{B}}
    \rangle\!\rangle\;,
\label{eq:res_quad}
\end{equation}
where I used independence of the baryon and the anti-baryon
distributions.  It should be noted that Eq.~\eqref{eq:res_quad}
exactly coincides with the formula derived in
Ref.~\cite{Kitazawa:2012at}.

Let us see how large the second term could be, and for this purpose,
I make use of an expression for the free baryon gas.  Then, I
numerically confirm that this second term is very close to the first
term at good precision; i.e.,
$\langle\!\langle\calN_B+\calN_{\bar{B}}\rangle\!\rangle\approx \chi_B^{(2)}$
within 1\% level at large $\snn$ and at most 5\% level at smaller
$\snn$ of a few GeV.\ \ I can then approximate $\chi_p^{(2)}$ as
$\chi_p^{(2)}\approx (1/2)\chi_B^{(2)}$.  This means that both
Eq.~\eqref{eq:res_quad} and the previous relation in the HRG model
eventually lead to the same answer; $\chi_p^{(2)}=(1/2)\chi_B^{(2)}$
after all, though they superficially look quite different from each
other.

I next proceed to the $n=3$ case.  Then, after some calculations, I
can arrive at:
\begin{equation}
 \chi_p^{(3)} = \frac{1}{8}\chi_B^{(3)} + \frac{3}{8}
  \bigl( \calchi_B^{(2)} - \calchi_{\bar{B}}^{(2)} \bigr)\;,
\label{eq:res_cubic}
\end{equation}
where I defined $\calchi_B^{(2)}$ and $\calchi_{\bar{B}}^{(2)}$ as
\begin{align}
 \calchi_B^{(2)} &\equiv \frac{1}{VT^3} \bigl(
  \langle\!\langle \calN_B^2\rangle\!\rangle
   - \langle\!\langle \calN_B\rangle\!\rangle^2 \bigr) \;,\\
 \calchi_{\bar{B}}^{(2)} &\equiv \frac{1}{VT^3} \bigl(
  \langle\!\langle \calN_{\bar{B}}^2\rangle\!\rangle
   - \langle\!\langle \calN_{\bar{B}}\rangle\!\rangle^2 \bigr)\;.
\end{align}
So, the ordinary quadratic fluctuation is given as
$\chi_B^{(2)}=\calchi_B^{(2)}+\calchi_{\bar{B}}^{(2)}$.  My result
above is again equivalent to the formula listed in
Ref.~\cite{Kitazawa:2012at}.  It is also easy to check that this
latter term in Eq.~\eqref{eq:res_cubic} gives the same answer as
$\chi_B^{(3)}$ within a few \% as long as the baryon distribution is
thermal.  Therefore, $\chi_p^{(3)}\approx (1/2)\chi_B^{(3)}$ follows.

\begin{figure}
 \includegraphics[width=\columnwidth]{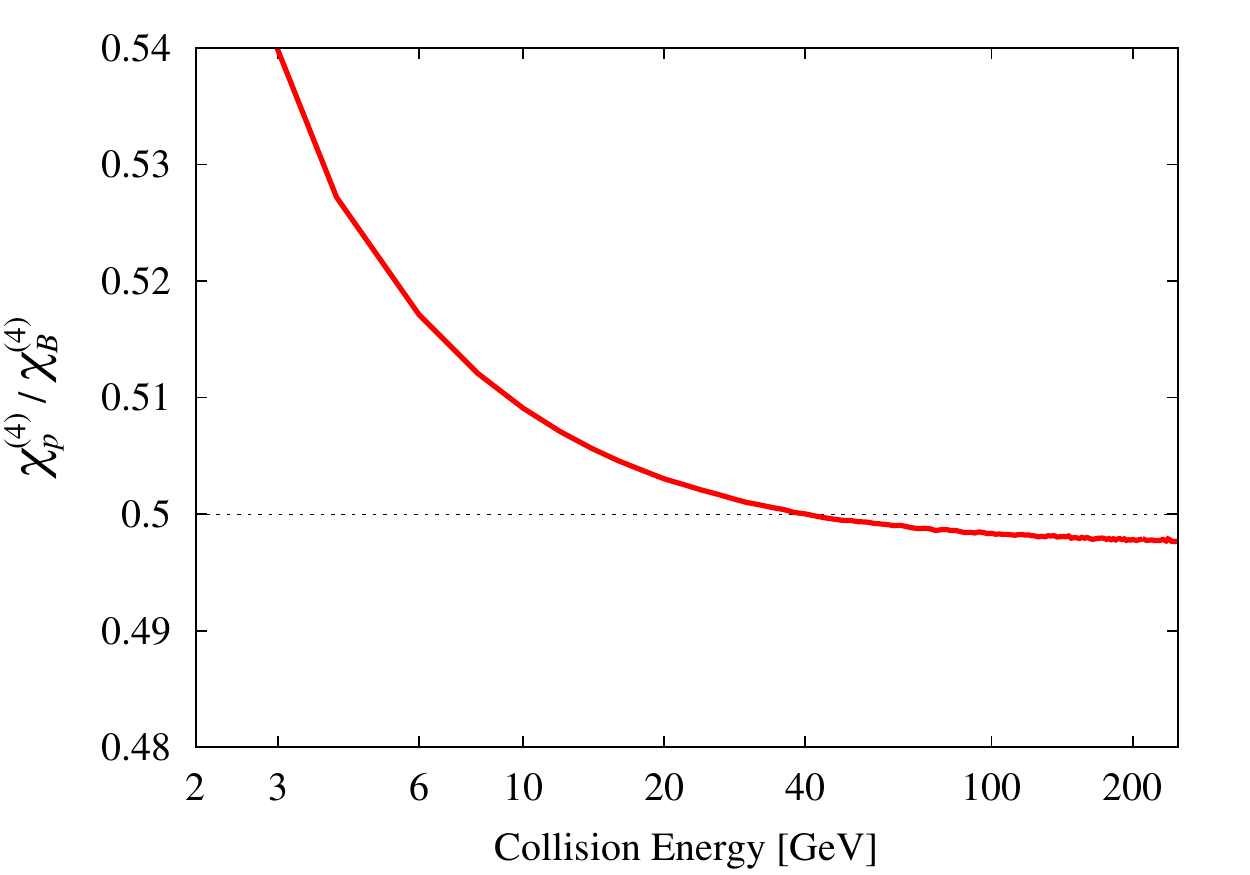}
 \caption{(Color online) Ratio of $\chi_p^{(4)}$ to $\chi_B^{(4)}$
   which is close to $1/2$ and the deviation from $1/2$ is less than
   10\% at small $\snn$.}
 \label{fig:isospin}
\end{figure}

Now I can make a guess that probably
$\chi_p^{(4)}\approx (1/2)\chi_B^{(4)}$ again and let us explicitly
make it sure.  In the same way I can write $\chi_p^{(4)}$ down as
\begin{equation}
 \begin{split}
 \chi_p^{(4)} &= \frac{1}{16}\chi_B^{(4)}
  + \frac{3}{8}\bigl( \calchi_B^{(3)} + \calchi_{\bar{B}}^{(3)} \bigr)\\
 & \qquad + \frac{3}{16} \chi_B^{(2)}
   - \frac{1}{8VT^3} \langle\!\langle \calN_B + \calN_{\bar{B}}
     \rangle\!\rangle \;.
 \end{split}
\end{equation}
This is indeed close to $(1/2)\chi_B^{(4)}$ but shows a deviation as
$\snn$ gets smaller.  I present our numerical results in
Fig.~\ref{fig:isospin}.  It is clear from Fig.~\ref{fig:isospin} that
$\chi_p^{(4)}\approx (1/2)\chi_B^{(4)}$ is the case as long as $\snn$
is sufficiently large, while it increases by about 10\% at smaller
$\snn$.  My conclusion is that, contrary to what is claimed in
Ref.~\cite{Kitazawa:2012at}, the isospin correlation does not help us
with explaining a suppression tendency in the kurtosis at smaller
$\snn$;  the effect is in a wrong direction.  In any case, the 10\%
correction is just too minor to account for almost 50\% suppression in
the experimental data as seen in Fig.~\ref{fig:kurtosis}.

Here I make a remark that I can easily give a general proof of
$\chi_p^{(n)}\approx (1/2)\chi_B^{(n)}$ if I can make the Boltzmann
approximation for the baryon distribution.  Therefore, in this sense,
the 10\% deviation seen in Fig.~\ref{fig:isospin} can be attributed to
the violation of the Boltzmann approximation that is quantified by the
deviation from the unity in Fig.~\ref{fig:kurtosis}, which is also of
the 10\% level.  The bottom line of my analysis is that I can safely
neglect the difference between the baryon number and the proton number
fluctuations.

\section{Summary}
\label{sec:summary}

I investigated the baryon number fluctuations using the hadron
resonance gas model and the mean-field model of nuclear matter.  I
found that the mean-field description yields fairly good results which
look quite consistent with the skewness and the kurtosis measured in
the beam-energy scan.

Because the mean-field approximation is based on the quasi-particle
treatment, in fact, it is not much different from the hadron resonance
gas model except for the interaction effects incorporated in terms of
the scalar and the vector mean-fields.  I numerically checked that
the kurtosis is suppressed at smaller collision energy (i.e., higher
baryon density) due to the vector mean-field that is directly coupled
to the baryon density.
I would emphasize that my main point is to draw attention to a
realistic possibility to interpret the BES data as an extrapolation
from nuclear matter, and not to make a serious comparison between
models and the experimental data.  To this end I need to take account
of canonicalness in a finite volume~\cite{Bzdak:2012an} and also
diffusion effects in rapidity subspace~\cite{Sakaida:2014pya}.

Finally, in the present study, I discussed the effects of isospin
correlations and reached a conclusion that such effects are only minor
such that I can ignore them in the first approximation.  Even in the
case of strong residual interactions that realize complete
randomization in isospin space, I found that the deviation from the
HRG prediction is at most 10\% at the smallest collision energy of a
few GeV.\ \  Therefore, for a semi-quantitative estimate, I can
simply identify the proton number fluctuations as (a half of) the
baryon number fluctuations.

In this paper I only mentioned on another possibility of flavor
mixing through the off-diagonal susceptibility: $\chi_{ud}$.  This
non-zero $\chi_{ud}$ arises from the pion dynamics, and so it is quite
non-trivial how we can relate $\chi_{ud}$ to the correlations purely
among the proton number $N_p$ and the neutron number $N_n$.  I am
now making progress in this direction in order to refine relationship
between $\chi_p^{(n)}$ and $\chi_B^{(n)}$.

Although the $\sigma$-$\omega$ model is one of the simplest methods to
capture the essential features of nuclear matter, it would be more
desirable to develop quantitative investigations by means of more
systematic approaches such as the Chiral Perturbation Theory.  It
would be definitely worth attempting the fully quantitative
comparisons for $S\sigma$ and $\kappa\sigma^2$ within the framework of
the Chiral Perturbation Theory and also more established Bruckner-type
calculations.  This is one of my future problems and the results
shall be reported in follow-ups hopefully soon.

\acknowledgments
The author thanks Masayuki~Asakawa and Masakiyo~Kitazawa for useful
discussions.  This work was supported by JSPS KAKENHI Grant Number
24740169.



\begin{thebibliography}{99}

\bibitem{BraunMunzinger:2008tz}
  P.~Braun-Munzinger and J.~Wambach,
  Rev.\ Mod.\ Phys.\  {\bf 81}, 1031 (2009).

\bibitem{Fukushima:2010bq}
  K.~Fukushima and T.~Hatsuda,
  Rept.\ Prog.\ Phys.\  {\bf 74}, 014001 (2011);
  K.~Fukushima,
  J.\ Phys.\ G {\bf 39}, 013101 (2012);
  K.~Fukushima and C.~Sasaki,
  Prog.\ Part.\ Nucl.\ Phys.\  {\bf 72}, 99 (2013).

\bibitem{Aarts:2013bla}
  G.~Aarts,
  PoS LATTICE {\bf 2012}, 017 (2012).

\bibitem{Fukuda:2013ada}
  R.~Fukuda, K.~Fukushima, T.~Hayata and Y.~Hidaka,
  Phys.\ Rev.\ D {\bf 89}, 014508 (2014).

\bibitem{Stephanov:1998dy}
  M.~A.~Stephanov, K.~Rajagopal and E.~V.~Shuryak,
  Phys.\ Rev.\ Lett.\  {\bf 81}, 4816 (1998);
  Phys.\ Rev.\ D {\bf 60}, 114028 (1999).

\bibitem{Fukushima:2008is}
  K.~Fukushima,
  Phys.\ Rev.\ D {\bf 78}, 114019 (2008).

\bibitem{Nakano:2004cd}
  E.~Nakano and T.~Tatsumi,
  Phys.\ Rev.\ D {\bf 71}, 114006 (2005).

\bibitem{Nickel:2009ke} 
  D.~Nickel,
  Phys.\ Rev.\ Lett.\  {\bf 103}, 072301 (2009).

\bibitem{Fukushima:2012mz}
  K.~Fukushima,
  Phys.\ Rev.\ D {\bf 86}, 054002 (2012).

\bibitem{Buballa:2014tba} 
  For a review, see; M.~Buballa and S.~Carignano,
  Prog.\ Part.\ Nucl.\ Phys.\  {\bf 81}, 39 (2015).

\bibitem{Adamczyk:2013dal} 
  L.~Adamczyk {\it et al.}  [STAR Collaboration],
  Phys.\ Rev.\ Lett.\  {\bf 112}, no. 3, 032302 (2014).

\bibitem{Asakawa:2009aj} 
  M.~Asakawa, S.~Ejiri and M.~Kitazawa,
  Phys.\ Rev.\ Lett.\  {\bf 103}, 262301 (2009).

\bibitem{Stephanov:2008qz} 
  M.~A.~Stephanov,
  Phys.\ Rev.\ Lett.\  {\bf 102}, 032301 (2009).

\bibitem{Cleymans:2005xv} 
  J.~Cleymans, H.~Oeschler, K.~Redlich and S.~Wheaton,
  Phys.\ Rev.\ C {\bf 73}, 034905 (2006).

\bibitem{Becattini:2005xt} 
  F.~Becattini, J.~Manninen and M.~Gazdzicki,
  Phys.\ Rev.\ C {\bf 73}, 044905 (2006).

\bibitem{Andronic:2008gu} 
  A.~Andronic, P.~Braun-Munzinger and J.~Stachel,
  Phys.\ Lett.\ B {\bf 673}, 142 (2009)
  [Erratum-ibid.\ B {\bf 678}, 516 (2009)].

\bibitem{Borsanyi:2011zz} 
  S.~Borsanyi, G.~Endrodi, Z.~Fodor, C.~Hoelbling, S.~Katz, S.~Krieg, C.~Ratti and K.~Szabo,
  J.\ Phys.\ Conf.\ Ser.\  {\bf 336}, 012019 (2011).

\bibitem{Karsch:2010ck} 
  F.~Karsch and K.~Redlich,
  Phys.\ Lett.\ B {\bf 695}, 136 (2011).

\bibitem{Stachel:2013zma} 
  J.~Stachel, A.~Andronic, P.~Braun-Munzinger and K.~Redlich,
  J.\ Phys.\ Conf.\ Ser.\  {\bf 509}, 012019 (2014).

\bibitem{Tatsumi:2011tt} 
  T.~Tatsumi, N.~Yasutake and T.~Maruyama,
  arXiv:1107.0804 [nucl-th].

\bibitem{Floerchinger:2012xd} 
  S.~Floerchinger and C.~Wetterich,
  Nucl.\ Phys.\ A {\bf 890-891}, 11 (2012).

\bibitem{Walecka:1974qa} 
  J.~D.~Walecka,
  Annals Phys.\  {\bf 83}, 491 (1974).

\bibitem{Drews:2013hha} 
  M.~Drews, T.~Hell, B.~Klein and W.~Weise,
  Phys.\ Rev.\ D {\bf 88}, no. 9, 096011 (2013).

\bibitem{Andronic:2009gj} 
  A.~Andronic, D.~Blaschke, P.~Braun-Munzinger, J.~Cleymans, K.~Fukushima, L.~D.~McLerran, H.~Oeschler and R.~D.~Pisarski {\it et al.},
  Nucl.\ Phys.\ A {\bf 837}, 65 (2010).

\bibitem{Fukushima:2009dx} 
  K.~Fukushima,
  Phys.\ Rev.\ D {\bf 79}, 074015 (2009).

\bibitem{Buballa:1996tm} 
  M.~Buballa,
  Nucl.\ Phys.\ A {\bf 611}, 393 (1996).

\bibitem{Blaizot:1980tw} 
  J.~P.~Blaizot,
  Phys.\ Rept.\  {\bf 64}, 171 (1980).

\bibitem{sasaki}
  T.~Sasaki, in private communications.

\bibitem{Chomaz:2004nw} 
  P.~Chomaz,
  nucl-ex/0410024.

\bibitem{Gavai:2005yk} 
  R.~V.~Gavai and S.~Gupta,
  Phys.\ Rev.\ D {\bf 73}, 014004 (2006);
  Phys.\ Rev.\ D {\bf 78}, 114503 (2008).

\bibitem{Mukherjee:2006hq} 
  S.~Mukherjee, M.~G.~Mustafa and R.~Ray,
  Phys.\ Rev.\ D {\bf 75}, 094015 (2007).

\bibitem{Cristoforetti:2010sn} 
  M.~Cristoforetti, T.~Hell, B.~Klein and W.~Weise,
  Phys.\ Rev.\ D {\bf 81}, 114017 (2010).

\bibitem{Fukushima:2003fw} 
  K.~Fukushima,
  Phys.\ Lett.\ B {\bf 591}, 277 (2004).

\bibitem{Megias:2004hj} 
  E.~Megias, E.~Ruiz Arriola and L.~L.~Salcedo,
  Phys.\ Rev.\ D {\bf 74}, 065005 (2006).

\bibitem{Kitazawa:2012at} 
  M.~Kitazawa and M.~Asakawa,
  Phys.\ Rev.\ C {\bf 85}, 021901 (2012);
  Phys.\ Rev.\ C {\bf 86}, 024904 (2012)
  [Erratum-ibid.\ C {\bf 86}, 069902 (2012)].

\bibitem{Bzdak:2012an} 
  A.~Bzdak, V.~Koch and V.~Skokov,
  Phys.\ Rev.\ C {\bf 87}, 014901 (2013).

\bibitem{Sakaida:2014pya} 
  M.~Sakaida, M.~Asakawa and M.~Kitazawa,
  Phys.\ Rev.\ C {\bf 90}, 064911 (2014).

\end{thebibliography}
\end{document}